\documentclass[a4paper]{article}

\usepackage[english]{babel}
\usepackage[utf8]{inputenc}
\usepackage{amsmath}
\usepackage{graphicx}
\usepackage[colorinlistoftodos]{todonotes}
\usepackage{authblk}
\usepackage{float}
\usepackage{xurl}
\usepackage{amsmath,amssymb,amsfonts}
\usepackage{algorithmic}
\usepackage{graphicx}
\usepackage{textcomp}
\usepackage{xcolor}
\usepackage{multirow}
\usepackage{lipsum}
\usepackage{subfigure}
\usepackage{svg}
\usepackage{atbegshi}

\def\thickhline{\noalign{\hrule height.8pt}}
\DeclareUnicodeCharacter{2212}{\ensuremath{-}}
\usepackage[backend=biber,style=numeric, sorting=none]{biblatex}
\usepackage{fancyhdr}
\newcounter{dummy}

\title{On the feasibility of laser satellite communications from the
	Martian surface}

\author[1]{Eva Fernandez Rodriguez}
\author[2]{Zachary C. Rowland}
\author[3]{Roderik A. Overzier}

\affil[1]{Netherlands Organisation for Applied Scientific Research (TNO), The Netherlands}
\affil[2]{TU Delft, Civil Engineering and Geosciences, The Netherlands}
\affil[3]{TNO \& Leiden Observatory, Leiden University, The Netherlands}

\date{}

\begin{document}
	\maketitle
	
	\begin{abstract}
		Free space optical (FSO) communication using lasers is a rapidly developing field in telecommunications that can offer advantages over traditional radio frequency technology. For example, optical laser links may allow transmissions at far higher data rates, require less operating power and smaller systems and have a smaller risk of interception. In recent years, FSO laser links have been demonstrated, tested or integrated in a range of environments and scenarios. These include FSO links for terrestrial communication, between ground stations and cube-sats in low Earth orbit, between ground and satellite in lunar orbit, as part of scientific or commercial space relay networks, and deep space communications beyond the moon. The possibility of FSO links from and to the surface of Mars could be a natural extension of these developments. In this paper we evaluate some effects of the Martian atmosphere on the propagation of optical communication links, with an emphasis on the impact of dust on the total link budget. We use the output of the Mars Climate Database to generate maps of the dust optical depth for a standard Mars climatology, as well as for a warm (dusty) atmosphere. These dust optical depths are then extrapolated to a wavelength of 1.55 $\mu$m, and translated into total slant path optical depths to calculate link budgets and availability statistics for a link between the surface and a satellite in a sun-synchronous orbit. The outcomes of this study are relevant to potential future missions to Mars that may require laser communications to or from its surface. For example, the results could be used to constrain the design of communication terminals suitable to the Mars environment, or to assess the link performance as a function of ground station location.
	\end{abstract}

\refstepcounter{dummy}
\label{dummyref}
	
	\section{Introduction}
	Free space optical (FSO) communication using lasers is a rapidly developing field in telecommunications that can offer advantages over traditional radio frequency technology. For example, optical laser links may allow transmissions at far higher data rates, require less operating power and smaller systems, have a far smaller risk of interception, and could also be used to relay information between antennas without a direct line of sight while maintaining low latency. In recent years, several technology demonstrators have tested FSO laser links in a range of environments and scenarios. These included FSO links for terrestrial communication, between ground stations and low earth orbit \cite[TeraByte Infrared Delivery (TBIRD);][]{Riesing23}, between ground and satellite in lunar orbit \cite[Lunar Laser Communication Demonstration (LLCD);][]{Boroson14}, as part of space relay networks between ground and, for example, the International Space Station \cite[Integrated Low-Earth Orbit LCRD User Modem and Amplifier Optical Communications Terminal (ILLUMA-T);][]{Robinson18} using the Laser Communications Relay Demonstration (LCRD) \cite{Israel17}, and deep space communications beyond the moon such as the Deep Space Optical Communication (DSOC) project \cite{Biswas17}. In the future, high-definition direct-to-Earth laser communications are planned for the Orion crew vehicle on the lunar surface \cite[Orion Artemis II Optical Communications System (O2O);][]{Robinson18}. 
	
	FSO links from and to the surface of Mars could be a natural extrapolation of these developments. In fact, in the 1980s optical links had already been envisioned for the future Mars Rovers \cite[e.g.][]{Sipes86,Annis87}. \cite{Annis87} presented a first review of parameters of interest to the optical channel modeling on Mars based on data derived from the Viking 1 and 2 and Mariner 9 missions. That paper anticipated the use of a Nd:YAG laser at 0.53 $\mu$m, for which ample optical imaging at $\approx$0.4--0.7 $\mu$m obtained by the Viking Orbiters and Landers could be used to provide initial assessments of atmospheric transmission and blurring. A summary of that work, followed by a small review of more recent work that provides much more detail will be presented in Sect. 2. 
	
	Although FSO laser communication from Mars has been considered before, previous studies were based on either limited atmospheric and climatological data \cite[e.g.][]{Sipes86,Annis87} or focused purely on link budget and modulation schemes \cite[e.g.][]{Booth18}. \textcite{Rowland22} explored possible network architectures that would allow communication from the Martian surface to orbit, including possible relay options during extreme dust storms. In this paper, we will continue on that theme by exploring optical channel modeling and link budgets under a variety of scenarios and dust conditions. 
	
	The structure of this paper is as follows. In Sect. ~\ref{sec:mcd} the data source used for the analysis is discussed, Sect. \ref{sec:methodology} outlines the methodology for evaluating the link budget performancer, Sect. \ref{sec:scenarios} describes the different scenarios considered in the study, and Sects. \ref{sec:results} and \ref{sec:conclusions} include the results and conclusions respectively.

	\section{Atmospheric modeling}
	\label{sec:mcd}
	\subsection{General description}
	
	An early global overview of atmospheric conditions and optical depth measurements was given by \cite{Annis87}, which is summarized here. 
	
	Due to the thinness of the Martian atmosphere, the scattering by atmospheric molecules is negligible (optical depth $\tau\approx0.002$). Molecular absorption is dominated by the same bands as on Earth, although they are weak overall due to the thinness of the atmosphere. CO$_2$ is relatively strong, and H$_2$O is relatively weak, in accordance with their atmospheric proportions (Table \ref{tab:atmosphere}).
	
	\begin{table}[H]
		\caption{\label{tab:atmosphere}Comparison of atmospheric compositions of Earth and Mars.} 
		\centering
		\begin{tabular}{lllll}
			\thickhline
			Molecule & Earth (wet) & Mars \\
			\hline
			N$_2$ & 75 \% & 2.7 \%\\
			O$_2$ & 20.1 \% & 0.13 \%\\ 
			CO$_2$ & 0.9 \% & 95.3 \%\\
			H$_2$O & 4 \% & 0.03 \%\\
			Ar & 0.9 \% & 1.6 \% \\
			CO & 0.1 ppm & 0.07 ppm\\
			\thickhline
		\end{tabular}
	\end{table}
	
	The attenuation on Mars is dominated by aerosol scattering in the form of dust, clouds, fog and haze. Taking all these effects together, the optical depth data derived from images toward the Sun and Phobos from the landing sites of Viking 1 (+22$^\circ$\ N.) and 2 (+48$^\circ$\ N.) showed $\tau\sim0.5-1.0$ over a large fraction of the year, which was linked to the existence of global dust storms. The minimum found was $\tau\approx0.2-0.3$, presumably due to local $\sim$5 $\mu$m dust particle haze, while peaks of $\tau=1-4$ lasting from several days to many tens of days are also present. Single day and strong seasonal variations were found linked to the North Polar Hood (NPH), a system of (white) clouds that develops from Summer to Spring. The NPH can extend down to +30-40$^\circ$ N., and is smaller or absent during dust storms. Local dust storms last about one day. They are found at the edge of the South Polar Cap, and at the lower S. latitudes especially at Perihelion. Global storms last tens of days, from Spring to Summer in the S. They are caused by solar heating that magnifies local dust storms. Storms seen by Viking 1 had $\tau\approx3-4$. 
	
	In-orbit imaging found a few percent of Mars covered in cloud or fog, and about 30 \% in haze. In the same season, the N. is generally cloudier than the S., and cloud coverage in the N. increases towards the higher latitudes. The thick H$_2$O clouds do not occur at high latitudes (N. and S.) during late Autumn/Winter, and not at mid latitudes in the S. during early Winter. The typical clouds have $\tau\sim0.05-3$, which is similar to Earth cirrus clouds. Optically thick clouds ($\tau\gtrsim1$) frequently occur in the form of (seasonal) H$_2$O and CO$_2$ clouds and polar hood haze, in the Tharsis Bulge on the NW. flank of large volcanoes, and in fields of cellular clouds S. of the Western Marineris Valley during Summer. Isolated clouds include lee clouds (downwind volcanoes), wave clouds not linked to surface features, and CO$_2$ cirrus ($\tau\sim0.05$). Optically thick hazes of dust or ice also occur. A diurnal variation of $\Delta\tau\sim0.2$ is associated with H$_2$O fog, which disappears during the hours after Sunrise. Thick fog can be found at the bottom of craters, mostly in the S. and in Spring/Summer. 
	
	Using data at 450, 670, 883 and 989 nm from the Imager for Mars Pathfinder, \cite{Smith99} found slight increases in the blue opacity during mornings, while the longer wavelength opacities were stable during the day. This was interpreted as having more small particles in the morning that Rayleigh scatter the blue light. Because smaller particles tend to be suspended in an atmosphere longer than larger particles, it was deemed unlikely that the increased blue opacity is due to smaller dust particles. However, the data is consistent with small water ice aerosols (sizes less than about 0.3 $\mu$m). $H_{2}O$ clouds above the dust layer have been seen by the Hubble Space Telescope (\cite{James96}). \cite{Smith99} conclude that these water ice particles are responsible for an optical depth increase of $\tau\simeq0.05-0.15$ at blue wavelengths, while they have no effect in the red.    
	
	Summarizing these results, the gaseous atmosphere has a negligible impact on the total transmission, while haze/dust or fog/clouds dominate the optical depth with background values of a few tenths under normal conditions to $\tau\gg1$ in the case of dust storms or thick clouds. Because clouds have limited occurrence, tend to be highly localized, and only add limited opacity long-ward of 1 $\mu$m, we focus here on the opacity contribution due to dust only.       
	
	\subsection{Mars Climate Database}
	
	The Mars climate database (MCD; \cite{MCD}) is a database of meteorological fields that describe the Martian atmosphere. These fields are derived from numerical simulations of the Martian atmosphere performed using the Mars planetary climate model (Mars PCM) general circulation model (GCM) and are validated using observational data.
	
	The observational data covers dust climatology of more than a decade of Martian years and is derived from measurements obtained through TES (Thermal Emission Spectrometer onboard the Mars Global Surveyor launched in 1996 \cite{TES}), THEMIS (THermal Emission Imaging System onboard the Mars Odyssey launched in 2001 \cite{THEMIS}), MCS (Mars Climate Sounder onboard the Mars Reconnaissance Orbiter launched in 2005), and MERs (Mars Exploration Rovers) instruments \cite{MCD}. 
	
	
	The MCD includes mean and statistical values of parameters that impact the performance of free space optical communications such as atmospheric temperature and pressure as well as dust, water vapor and ice content values. Dust loading of the atmosphere is particularly relevant for FSO performance due to its impact on the optical depth of the atmosphere.
	
	The MCD can provide information for a series of synthetic dust scenarios: a standard Mars year (climatology), cold (i.e. extremely clear, low dust), warm (i.e. dusty atmosphere) and dust storm (i.e. Mars during a global dust storm) \cite{MCD}. All scenarios include the option of solar minimum, average or maximum Extreme UltraViolet (EUV) forcings. In this regard, the MCD provides daily and monthly mean values of parameters relevant for FSO communications assessment including the mean dust column visible optical depth above surface. This value is provided as a measurement from the local surface to the top of the atmosphere at a wavelength of 0.67 \(\mu\)m.
	
	\subsection{Optical turbulence}
	
	It is well known that a laser beam propagating through the Earth's atmosphere experiences scintillation as a consequence of optical turbulence. This turbulence is due to minute gradients in the local air temperature that affect the density and hence the optical refractive index. In this section we will show that for most practical purposes, optical turbulence is negligible on Mars (\cite{Annis87,Chide24}). 
	
	In the so-called Rytov approximation, scintillation of a laser beam is described by (\cite{Andrews06}):
	
	\begin{equation}
	\sigma_I = 1.23 C_n^2 k^{7/6}x^{11/6},
	\end{equation}
	
	\noindent
	where $k=2\pi/\lambda$, $x$ is the length of the propagation path and $C_n^2$ the refractive index structure constant, an expression for the strength of the turbulence. Values for $C_n^2$ on Earth at the surface typically range from $\sim10^{-16}$ to $10^{-12}$ m$^{-2/3}$, corresponding to very low to strong turbulence. For the investigation presented in this paper, an important question concerns the value of $C_n^2$ on Mars. 
	
	The refractive index of the Martian atmosphere is much smaller than that of Earth. Taking into account the differences in atmospheric pressure and temperature at the surface between Earth (1000 hPa and 293 K) and Mars (610 Pa and 240 K), as well as the different atmospheric compositions, the refractive index on Mars is about 80$\times$ lower than that on Earth. As a result, the value for $C_n^2$ is substantially lower as well. For zero humidity, a common expression is (\cite{Wu21}):
	
	\begin{eqnarray}
	C_n^2=aL^{4/3}\left(C_1\frac{P}{T^2}\right)^2\left(\frac{\partial\theta}{\partial h}\right)^2,\\
	\theta=T\left(\frac{P_0}{P}\right)^{R/C_p}.
	\end{eqnarray}
	
	\noindent
	Here $a=2.8$, $L$ is the outer scale of turbulence, $P_0$ is a reference pressure at the surface, $R/C_p$ is the ratio of the gas constant to the molar isobaric heat capacity, $\theta$ is the potential temperature and $h$ is altitude.   
	The constant $C_1$ has a value of $\approx7.9\times10^{-5}$ for air, and $\approx1.31\times10^{-4}$ for the (dry) Martian atmosphere (\cite{Ho20}). The ratio $R/C_p$ is 0.284 for Earth and 0.257 for Mars (\cite{Petrosyan11}).

	\begin{eqnarray}
	C_T^2=aL^{4/3}\left(\frac{\partial\theta}{\partial h}\right)^2,
	\end{eqnarray}
	
	\noindent
	and assuming that $C_T^2(\mathrm{Mars})\lesssim C_T^2(\mathrm{Earth})$ (see \cite{Annis87,Ho20}), we expect that 
	$C_n^2(\mathrm{Mars})\lesssim2\times10^{-4} C_n^2(\mathrm{Earth})$. This implies optical turbulence strengths on Mars of $\simeq10^{-20}-10^{-16}$ m$^{-2/3}$ for the full range of $C_n^2$ values experienced on Earth. 
	We therefore neglect the optical turbulence in this analysis.  

	\section{Methodology}\label{sec:methodology}
	The goal of this paper is to understand the conditions and parameters that would enable FSO communications through the Mars atmosphere. We consider two main scenarios. The first scenario assumes an FSO link between a ground station located on the Martian surface and an orbiting satellite. 
	In this scenario we perform a planet-wide link budget analysis under a set of climatology conditions retrieved from the MCD. The second scenario we will consider assumes an FSO link between a ground station at a fixed location and various short-range communications across the surface or to airborne vehicles. The scenarios are described in more detail in Sect. \ref{sec:scenarios}.   
	
	\subsection{Dust opacity at 1.55 $\mu$m}
	
	In this paper we will assume that the broad-band dust opacity as determined at wavelengths shorter than 1 $\mu$m can be extrapolated to the FSO wavelength of 1.55 $\mu$m. We convert the optical depth results at 0.67 $\mu$m provided by the MCD to 1.55 \(\mu\)m following the Angstrom power law:
	
	\begin{equation}
	\label{eqn:angstrom}
	\frac{\tau_{1.55}}{\tau_{0.67}} = \left(\frac{0.67}{1.55}\right)^{\alpha}. 
	\end{equation}
	
	We use an Angstrom exponent, $\alpha$, of \(−0.082\), based on the average value for sol 1450 from \cite{GL084407}, such that $\tau_{1.55}\simeq1.1\times\tau_{0.67}$. In \cite{GL084407} values ranging from \(-0.16\) to \(-0.036\) are also cited, however this variation has only a small effect on the calculation of \(\tau\) due to the similarity between the two wavelengths. Therefore the specific choice of $\alpha$ is not considered critical for the results presented in this paper.
	
	\subsection{Link budget}\label{subsec:linkbudget}
	
	The Martian atmosphere is very thin and it presents a much smaller volume than the atmosphere on Earth. However, the properties of the atmosphere are modified by highly fluctuating amounts of suspended dust carried by winds \cite{dust_clima}, establishing dust as one of the main drivers of Martian climatology. The characteristics of Martian climatology are incorporated into the calculation of atmospheric losses in the link budget, with dust optical losses identified as the main contributor to these losses.
	
	\begin{equation}
	\label{eqn:link_budget}
	P_\mathrm{rx} = P_\mathrm{tx} + G_\mathrm{tx} + G_\mathrm{rx} - L_\mathrm{fs} - L_\mathrm{atm}, 
	\end{equation}
	
	\begin{equation}
	\label{eqn:link_margin}
	M = P_\mathrm{rx} - P_\mathrm{req}.
	\end{equation}
	
	The link budget calculation used in this paper is given by Eq.~\ref{eqn:link_budget}. This is a simplification of the standard FSO scenario, as we will only consider the primary gains and losses of a hypothetical system with $P_\mathrm{rx}$ and $P_\mathrm{tx}$ being the received and transmitted powers, $G_\mathrm{rx}$ and $G_\mathrm{tx}$ the received and transmitted gains, and $L_\mathrm{fs}$ and $L_\mathrm{atm}$ the free space and atmospheric losses. Eq. \ref{eqn:link_margin} represents the link margin, $M$, given the sensitivity of the receiver, as expressed by the difference between the power received and the minimum power required for transfer of the signal.

	\subsubsection{Loss contributions}\label{subsubsec:losscontributions}
	In scenarios where the transmitter and receiver are separated by long distances, one of the primary losses to consider in the link budget calculation is the FSO path loss, 
	
	\begin{equation}
	\label{eqn:free_space}
	L_\mathrm{fs} = -10 \log_{10}\left(\frac{\lambda}{4 \pi L}\right)^2,
	\end{equation}
	
	\noindent
	where $\lambda$ is the laser wavelength and $L$ the (slant) range between transmitter and receiver.
	
	Atmospheric attenuation is also an important source of losses. As described in Sect. \ref{sec:mcd}, the dust optical depth is typically the dominant factor of attenuation on Mars. The atmospheric loss in terms of the optical depth $\tau$ is given by
	
	\begin{equation}
	\label{eqn:atm_losses}
	L_\mathrm{atm} = -10 \log_{10} e^{-\tau},
	\end{equation}
	
	where $\tau$ represents the (wavelength dependent) atmospheric attenuation coefficient, $\alpha$, integrated along the path of transmission:
	
	\begin{equation}
	\label{eqn:opticaldepth}
	\tau = \int_{L_1}^{L_2} \alpha(l)\, dl.
	\end{equation}
	
	In this paper, we will use the optical depths determined for different levels of dust activity on Mars as extracted from the MCD. Further details on the optical depth profiles used are given in Sect. \ref{sec:optical_depth_param}. 
	
	\subsubsection{Gain contributions}\label{subsubsec:gaincontributions}
	
	The antenna transmission gain is typically calculated from the divergence angle of the transmitted beam $\theta$ as follows
	
	\begin{equation}
	\label{eqn:Gtx}
	G_\mathrm{tx} = 10 \log_{10} \left(\frac{8}{\theta^2}\right),
	\end{equation}
	
	while the receiver antenna gain is defined by its aperture size $D$:
	
	\begin{equation}
	\label{eqn:Grx}
	G_\mathrm{rx} = 10 \log_{10} \left(\frac{\pi D}{\lambda}\right)^2.
	\end{equation}
	
	\subsection{Optical depth profile parametrisation}
	\label{sec:optical_depth_param}
	
	Dust optical depth is only provided by the MCD in the form of the integrated value of $\tau$ for a column from the surface of Mars to the top of the atmosphere. Therefore, it cannot be readily applied to slant paths to an orbiting satellite, or to shorter paths between a transmitter and receiver closer to the surface.
	
	\begin{figure*}
		\centering
		\includegraphics[width=0.7\textwidth,keepaspectratio]{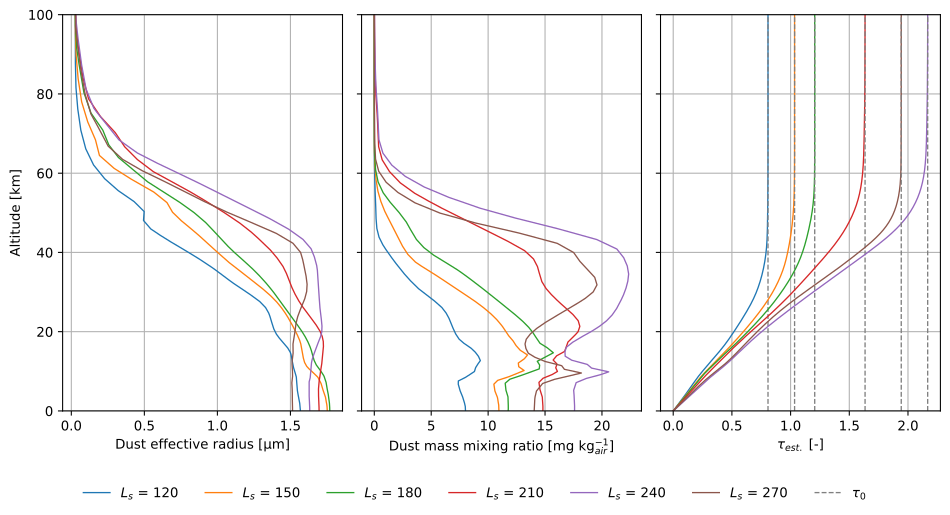}
		\caption{For dates ($L_s$) in months 5 - 10 at the Curiosity landing site (4.5895°S 137.4417°E) in the "warm" scenario of the MCD, from 0--100 km altitude, at local solar time 12h: (left) dust effective radius; (middle) dust mass mixing ratio; (right) estimated optical depth, $\tau_\mathrm{est}$.
		}
		\label{fig:mixingratio}
	\end{figure*}
	
	For a more flexible analysis of possible optical link paths, we have made a parametrisation of the vertical dust profile through the atmosphere. We assume that the dust mass mixing ratio profile as retrieved from the MCD correlates exactly with the vertical optical depth profile. Then, a cumulative $\tau$ profile can be estimated by scaling it with the normalised mass mixing ratio profile. Fig. \ref{fig:mixingratio} shows the dust effective radius and mass mixing ratios for dates (defined by solar longitude, $L_s$) in months 5 - 10 (months being defined as covering 30° of $L_s$ in the MCD) in the "warm" scenario of the MCD, as well as the cumulative value of optical depth, $\tau_\mathrm{est.}$, from the surface to 100 km as derived from the monthly mean dust column visible optical depth from the MCD at the Curiosity landing site, using the mass mixing ratio method.
	
	As can be seen in Fig. \ref{fig:mixingratio}, while the effective radius of the dust is fairly constant on some dates up to the thickest part of the atmosphere, this is not always the case. There will thus be wavelength-dependent effects on the optical depths that change with height. Currently, this is reserved for future work.
	
	\subsection{Slant path calculation}
	\label{sec:slant_path_calc}
	
	This paper considers two methods to calculate the slant path optical depth from the vertical profiles we extract from the MCD. We will show below that the results obtained using these two methods are quite comparable. 
	
	\subsubsection{Through air-mass scaling} \label{sec:slant_path_airmass}
	
	First, we can simply make the approximation that the normal to slant-path optical depths are related by the air mass factor. This ratio is given by Eq. \ref{eqn:slant_path}, where $\theta$ corresponds to the slant angle:
	
	\begin{equation}
	\label{eqn:slant_path}
	\frac{1}{\cos(\theta)}.
	\end{equation}
	
	This approximation is valid for angles of $\theta$ that are not too large (\cite{NIJEGORODOV199657}). Note that we define slant angle as the angle the transmitter (e.g. the satellite) makes with the zenith directly overhead the ground station.
	
	\subsubsection{Through optical depth profile parametrisation}
	\label{sec:slant_path_opticaldepth}
	
\begin{figure*}
		\centering
		\includegraphics[width=0.7\textwidth,keepaspectratio]{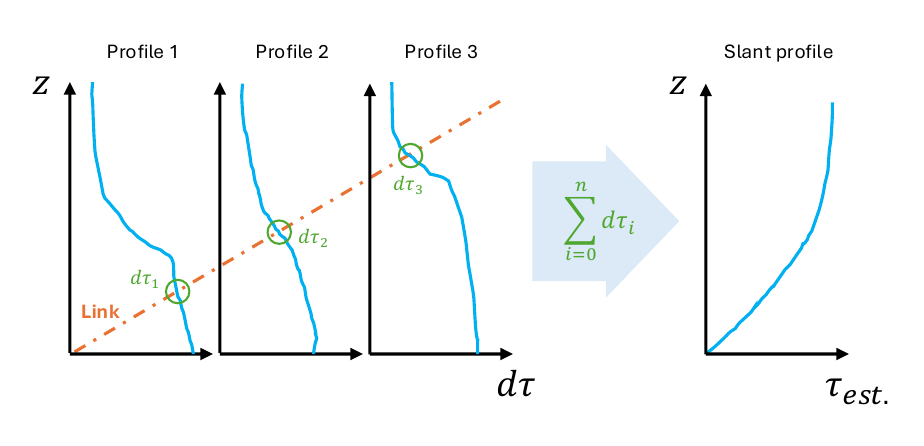}
		\caption{Illustration of the algorithm in Eq. \ref{eqn:tau_est_calc} used to calculate optical depths along slant paths through the atmosphere using the MCD data.}
		\label{fig:mixingratio_diagram}
	\end{figure*}

	The (cumulative) vertical optical depth profiles that we have obtained, as described in Sect. \ref{sec:optical_depth_param}, can also be used to calculate optical depths along a slant path, as illustrated in Fig. \ref{fig:mixingratio_diagram}. Our method consists of the following: $n$ three-dimensional spatial coordinates are calculated along a path from the ground station to the transmitter. At each altitude, the slant path intersects a specific vertical profile of optical depth, as calculated using the dust mass mixing ratio method described above. In Fig. \ref{fig:mixingratio_diagram} these points are indicated by ``profile 1", ``profile 2", ``profile 3", etc. At the altitude where the link intercepts the profile we take the vertical resolution used to calculate $d\tau$ for that profile and scale it using the correction factor in Eq. \ref{eqn:slant_path}, then recalculate $d\tau_i$ for that part of the link path based on the longer, slant increment resolution. We then sum all the values of $d\tau_i$ to calculate an estimated slant optical depth, $\tau_\mathrm{est.}$:
	
	\begin{equation}
	\label{eqn:tau_est_calc}
	\tau_\mathrm{est.} = \sum_{i=0}^{n} d\tau_i.
	\end{equation}

	\section{Scenarios}\label{sec:scenarios}
	The scenarios considered in this study were developed after a comprehensive assessment of both past and current Mars missions, paired with a forward-thinking consideration of the unique obstacles that a human settlement on the planet may encounter. 
	
	Having in mind the needs of future missions and human settlements on Mars, the definition of scenarios has been divided into two categories: atmospheric dust and operational scenarios. The atmospheric dust scenarios explore the impact of different dust concentrations on optical communications capabilities considering a satellite transmitting to a fixed ground station. The operational scenarios consider different configurations of transmitter and receiver for a single, dusty scenario.
	
	\subsection{Atmospheric dust scenarios}
	
	Atmospheric dust scenario results are determined for monthly averages of months 5 - 10 for the entire surface of Mars. Months 5 - 10 are chosen to cover the relatively clean northern summer to autumn months (5 - 7) and the dustier northern winter months (8 - 10), thus giving a representation of a variety of conditions on Mars.
	
	Values of monthly mean dust column visible optical depth (here labelled $\tau_0$, to specify the vertical column value) are extracted for a grid of 3.75$^\circ$\ latitude $\times$ 5.625$^\circ$\ longitude cells (the resolution of the MCD). We then calculated the maximum possible slant angle before the link margin (eq. \ref{eqn:link_margin}) falls below 3 dB for $\tau_\mathrm{est.}$ calculated using the correction in eq. \ref{eqn:slant_path}. The results consider a configuration where a satellite (transmitter) with an orbit altitude of 350 km above the Martian aeroid sends optical data to a ground station (receiver) on the surface of Mars.
	
	We also calculate statistics regarding how frequently the link margin of 3 dB is exceeded, both spatially across the planet (see Fig. \ref{fig:hist_global}) and in time for specific sites of interest (see Fig. \ref{fig:hist_sites}). The specific sites considered were:\\ 
	(1) the Phoenix landing site (68.2188$^\circ$ N, 125.7492$^\circ$ W, Green Valley, Vastitas Borealis), as it lies in the northern polar region and provides an example of a site in a relatively clean (non-dusty) region of Mars,\\
	(2) the Hellas Planitia (42.4$^\circ$ S, 70.5$^\circ$ E), a plain within the large Hellas impact crater which is associated with consistently high dust concentrations throughout the year,\\
	(3) the Curiosity landing site (4.5895$^\circ$ S, 137.4417$^\circ$ E, Bradbury Landing, Gale crater) as it lies on the boundary between the dusty northern plains and the southern highlands.
	
	The two different atmospheric scenarios analysed in this paper differ by their dust conditions. The dust condition characteristics are represented in the MCD through the ``climatology'' and ``warm'' scenarios described below.
	
	\begin{enumerate}
		\item \textbf{Climatology:} the dust distributions on the Mars PCM are reconstructed from observations from Mars years 24 - 35, resulting in a "standard year" climatology (i.e. no planet-encircling global dust storm). This scenario is for average solar EUV conditions.
		\item \textbf{Warm:} this corresponds to "dusty atmosphere" conditions where dust opacity at a given location is set to the maximum observed and further increased by 50\%, except during a global dust storm. This scenario also assumes a thermosphere under  maximum solar EUV conditions.
	\end{enumerate}
	
	The MCD also offers a ``global dust storm'' scenario, but this was not considered in this study. The dust optical depths are generally so large in this scenario that the link margin threshold is never met across almost the entire planet, even when the satellite is at zenith. Alternative possibilities for optical communications under such extreme conditions could be an interesting topic for future research.
	
	\subsection{Operational scenarios}\label{subsec:commsscenarios}
	
	The operational scenarios include different possibilities for links between a ground station on the Martian surface and transmitters flying overhead at different altitudes. This is done to observe the performance of the link in three possible mission scenarios, conveniently spaced in altitude.
	
	We apply the two slant path models described in Sect. \ref{sec:slant_path_calc} to calculate $\tau$ at each slant angle. In the case of the air-mass scaling model (Sect. \ref{sec:slant_path_airmass}), we calculate the initial vertical column value of $\tau_0$ at the altitude considered in each scenario by applying the optical depth profile parameterisation described in Sect. \ref{sec:optical_depth_param} to the entire atmospheric profile and then sum only the increments of $d\tau$ up to the altitude of the transmitter, to get the column optical depth for the dust from the ground station to the transmitter at zenith.
	
	In the case of the optical depth profile parameterisation of the slant path (\ref{sec:slant_path_opticaldepth}), the increments of $d\tau$ are calculated as the link intercepts each profile.
	
	\subsubsection{Satellite to ground}\label{subsubsec:sat2ground}
	
	Sun-synchronous orbits with heights between 300 and 400 km have been used in different missions such as Mars Global Surveyor (MGS), Mars Odyssey, and Mars Reconnaissance Orbiter (MRO). In this orbit type, the orbital plane precesses with nearly the same period as the planet's solar orbit which can be beneficial when the instruments onboard the satellite depend on a certain angle of solar illumination on the surface \cite{NASA_orbits}.
	
	Foreseeing the need of high throughput to download data from instruments onboard satellites in Martian orbits, an architecture using optical communications can be considered. In this scenario, the optical link is set between the surface of Mars and a satellite orbiting the planet in a sun-synchronous orbit (SSO) with an altitude of 350 km above the Mars aeroid where the satellite passes directly over the ground station on the surface of Mars during its orbit.

	\subsubsection{Balloon to ground}\label{subsubsec:balloon2ground}
	
	High-altitude balloons could be utilised for optical communications on Mars, as they are capable of flying above the densest part of the atmosphere and a large fraction of the depth of a relatively dusty atmosphere, reducing attenuation in the link budget. The concept of using a balloon to act as a rely between a ground station and a satellite on Mars was explored in \cite{Rowland22}. In this paper we consider the scenario of a ground station receiving from a balloon operating at 18 km above the Mars areoid, identified in \cite{Rowland22} as being an equivalent operating altitude (in terms of atmospheric density) to the maximum altitude of stratospheric balloons on Earth.

	\subsubsection{Helicopter to ground}\label{subsubsec:heli2ground}
	
	NASA's Ingenuity Mars helicopter demonstrated the feasibility of helicopters as a platform for planetary exploration, by operating in the extreme conditions of the Martian atmosphere (low density, higher radiation environment, limited power supply, etc.) and even performing operational flights as a scout for the Perseverance rover (\cite{ingenuity_ops}). In a typical flight on Mars, Ingenuity would reach a height of 10 metres travelling a horizontal distance of a few hundred metres. On the 5th of October 2023, Ingenuity achieved its maximum height of 24 metres \cite{Ingenuity_numbers}.
	
	The values reached by Ingenuity serve as a reference for feasibility considerations in future missions to Mars. In this study, we consider a helicopter travelling at 10 metres altitude, to measure the performance of the FSO link with a moving object in the densest part of the atmosphere.

	\section{Results and discussion}\label{sec:results}
	This section includes the results obtained for the different scenarios previously defined in Sect. \ref{sec:scenarios}. Table \ref{tab:param_general} includes the values for the common parameters when calculating the link budget for the different configurations, with the exception of the receiver and transmitter efficiencies that were both set to 0.8.  
	
	\begin{table}[H]
		\caption{\label{tab:param_general} Common parameters for the link budget scenarios.} 
		\centering
		\begin{tabular}{lllll}
			\thickhline
			Parameter & Value \\
			\hline
			Wavelength & 1550 $\mu$m\\
			Receiver antenna diameter & 800 mm\\
			Transmitter beam divergence & 380 $\mu$rad\\
			Transmitted power & 200 mW\\
			Required power & -35.5 dBm\\
			\thickhline
		\end{tabular}
	\end{table}
	
	For the receiver part of the system, we use values from the Optical Ground Station (OGS) at a TNO location in The Hague, Netherlands. We use the receiver antenna diameter for this OGS (800 mm) from \cite{TNOOGS}. Small satellite requirements were considered for the transmitter. This is reflected in the relatively low transmitted power of 200 mW set for the experiments as well as the transmitted beam divergence, inspired by the corresponding on-orbit value on the TBIRD laser communications (a CubeSat based mission) \cite{Riesing23}.

	\subsection{Atmospheric dust scenarios}
	
\begin{figure*}
		\centering
		\includegraphics[width=0.55\textheight,keepaspectratio]{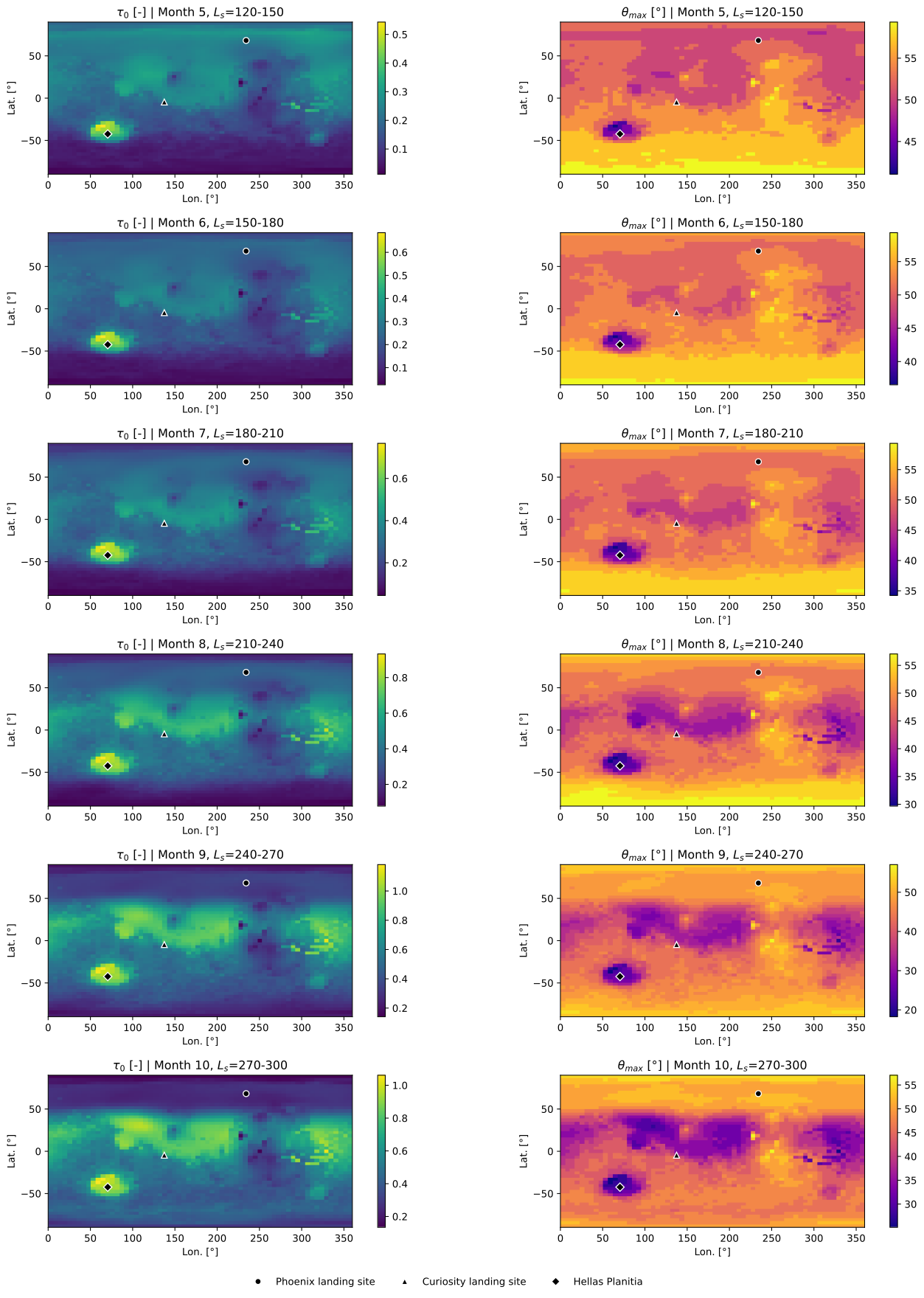}
		\caption{Results for the ``climatology" scenario from months 5-10. {\it Left panel:} MCD monthly mean dust column visible optical depth above the surface ($\tau_0$). {\it Right panel:} Maximum slant angle allowed between ground station and satellite assuming a minimum acceptable margin of 3 dB ($\theta_{max}$), calculated using the air mass scaling model (Sect. \ref{sec:slant_path_airmass}). The Hellas Planitia and the Phoenix and Curiosity landing sites are marked on the map.}
		\label{fig:map_climatology}
	\end{figure*}

\begin{figure*}
		\centering
		\includegraphics[width=0.55\textheight,keepaspectratio]{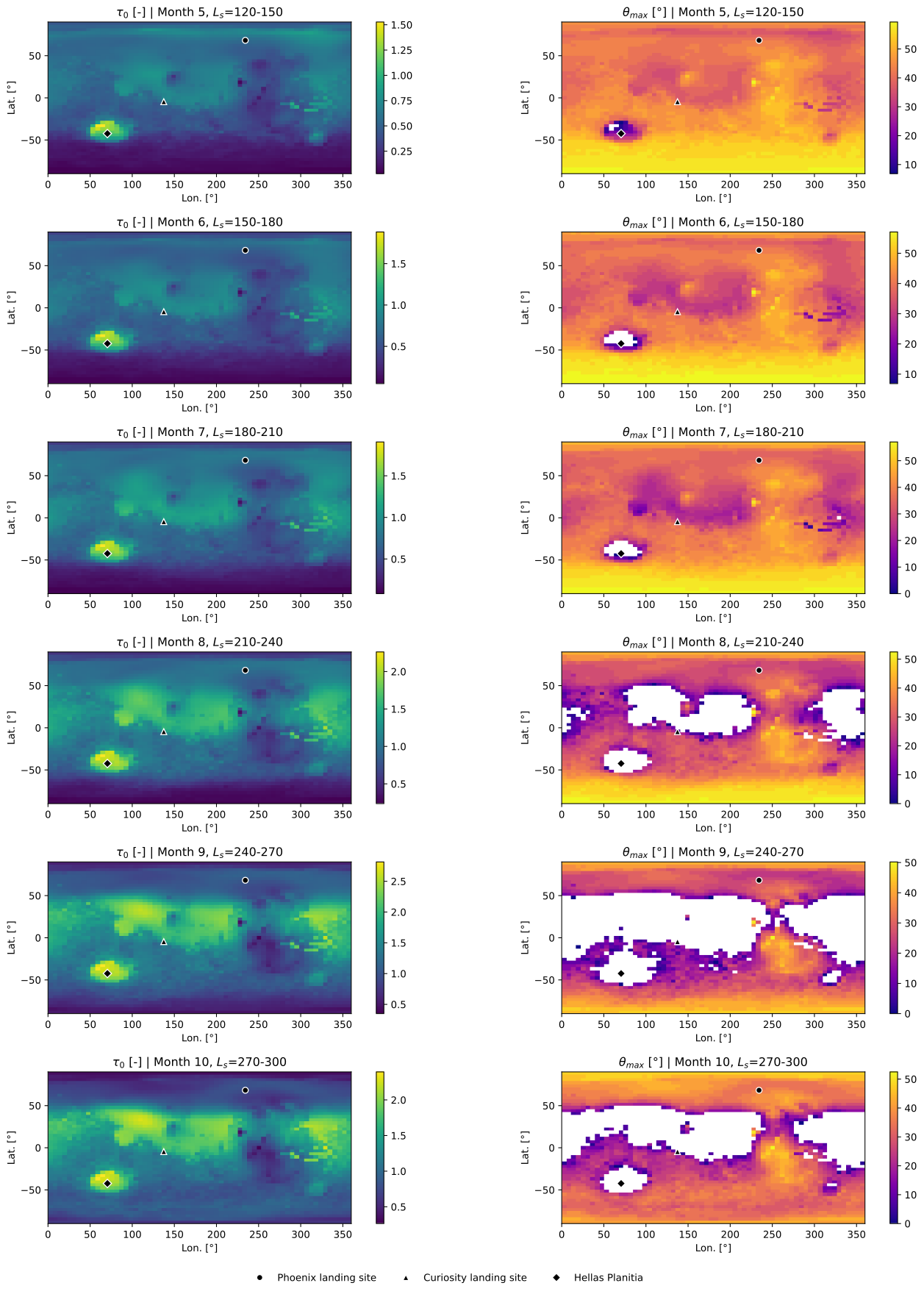}
		\caption{Results for the ``warm" scenario from months 5-10. {\it Left panel:} MCD monthly mean dust column visible optical depth above surface ($\tau_0$). {\it Right panel:} Maximum slant angle allowed between ground station and satellite assuming a minimum acceptable margin of 3 dB ($\theta_{max}$), calculated using the air mass scaling model (Sect. \ref{sec:slant_path_airmass}. White cells indicate areas where the link margin is exceeded even when the satellite is directly overhead (i.e., $\theta=0^\circ$). The Hellas Planitia and the Phoenix and Curiosity landing sites are marked on the map.}
		\label{fig:map_warm}
	\end{figure*}

	In Figs. \ref{fig:map_climatology} and \ref{fig:map_warm} (left columns) we show the monthly mean dust column optical depth ($\tau_0$) as retrieved from the MCD, for the ``climatology'' (Fig. \ref{fig:map_climatology}) and ``warm'' scenarios (Fig. \ref{fig:map_warm}). In the right columns of the corresponding figures we show the maximum slant angles achieved by a satellite in SSO transmitting to a ground station located at each grid cell on the map ($\theta_\mathrm{max}$).
	
	The maximum slant angle is calculated as angle beyond which the link margin exceeds a maximum permissible value of 3 dB, calculated using the air-mass model described in Sect. \ref{sec:slant_path_airmass}. A maximum slant path angle close to 0$^\circ$ implies that the optical link budget can only be closed when the satellite is directly over the ground station (i.e. at zenith). Maximum slant path angles closer to 90$^\circ$ imply that optical transmission is feasible even when passing through the thickest possible atmosphere, owing to the low dust optical depths in a certain region and month.
	
	In both scenarios it is possible to distinguish areas of high dust optical depth that align with regions with depressions in the terrain \cite{topmars} where the dust could be accumulating. During the ``climatology" scenario, optical communications are possible across the entire planet, albeit with variations in $\theta_\mathrm{max}$.
	
	In the ``warm'' scenario the values of dust optical depth are typically higher than in the ``climatology'' scenario. The white grid cells in Fig. \ref{fig:map_warm} indicate regions in which the link budget cannot be closed even with a satellite at zenith, making FSO communications impossible at that location in that month. This situation primarily occurs during the winter months in the northern plains when the dust concentration is highest, as well as in Hellas Planitia throughout the year.
	
	In the polar regions, there is a seasonal shift for both the ``climatology'' and ``warm'' scenarios but the optical depth remains low throughout the year. In the northern lowlands, dust tends to build up over the course of the 6 months considered, as the northern hemisphere transitions from summer to winter. In both scenarios, the mountainous Tharsis Rise region remains largely free of dust and consequentially the link margin remains low, allowing for relatively large $\theta_\mathrm{max}$. 
	
\begin{figure*}
		\centering
		\includegraphics[width=0.8\textwidth,keepaspectratio]{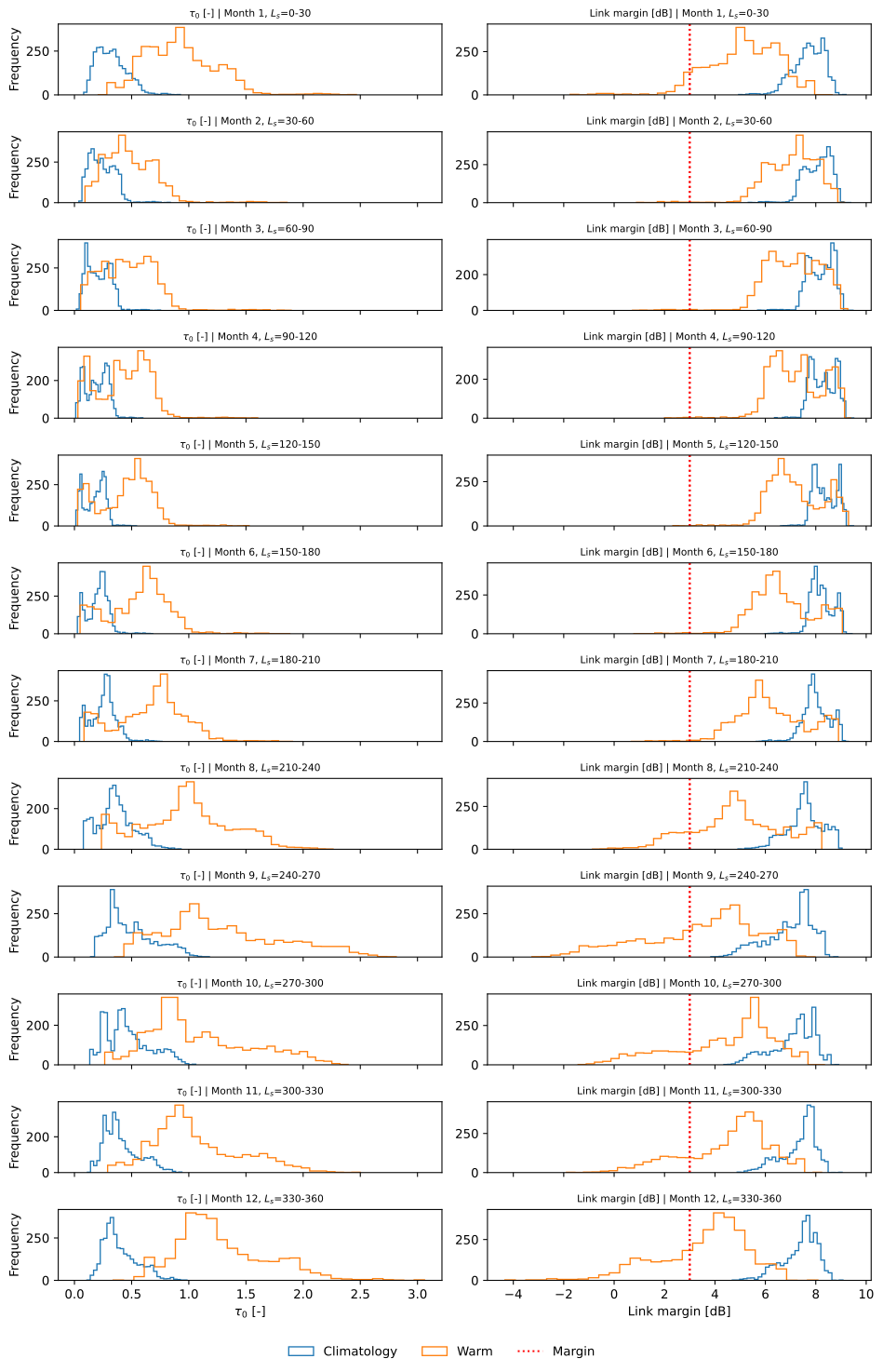}
		\caption{Frequency distributions of monthly mean dust column visible optical depth
			above surface, $\tau_0$, (left) and link margin (right), including both the ``climatology" (blue) and ``warm" (orange) scenarios. The red dashed line marks the threshold of 3 dB margin on the link budget needed to allow communications between transmitter and receiver.}
		\label{fig:hist_global}
	\end{figure*}

\begin{figure*}
		\centering
		\includegraphics[width=0.7\textwidth,height=\textheight,keepaspectratio]{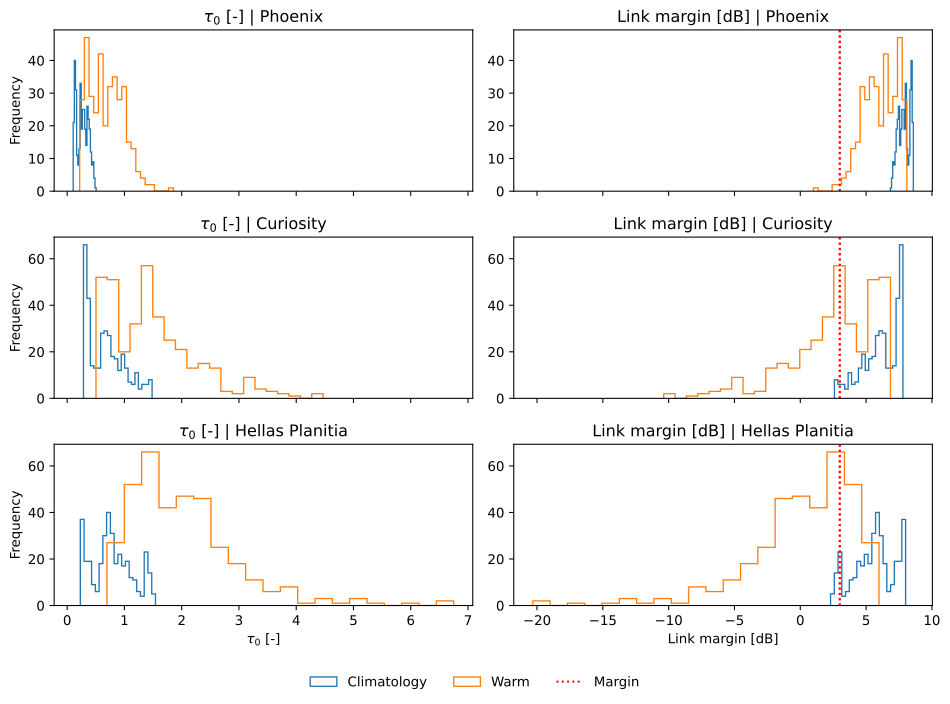}
		\caption{Frequency distributions of daily mean dust column visible optical depth above
			surface, $\tau_0$, (left) and link margin (right) for the Phoenix (top panels) and Curiosity (middle panels) landing sites, and Hellas Planitia (bottom panels) for one year duration. The results are given for both the ``climatology" (blue) and ``warm" (orange) scenarios. The red dashed line marks the threshold of 3 dB margin on the link budget needed to allow communications between transmitter and receiver.}
		\label{fig:hist_sites}
	\end{figure*}

	Figs. \ref{fig:hist_global} and \ref{fig:hist_sites} present the information from the maps shown in Figs. \ref{fig:map_climatology} and \ref{fig:map_warm} in the form of frequency distributions for the entire planet and at specific sites of interest, respectively, over an entire year (considering only $\tau$ at zenith, $\tau_0$). The $\tau_0$ distributions correlate well with the link margin values throughout the year, save for the additional (small) effect of the altitude of the ground station causing an additional variation of the FSO path loss (Eq. \ref{eqn:free_space}).
	
	When $\tau_0$ is low, the received power calculated in the link budget is relatively high, resulting in link margins over the 3 dB threshold. These margins indicate that optical communications are feasible. However, when $\tau_0$ increases, the received power becomes attenuated and the margins fall below the threshold for feasible communications.
	
	In Fig. \ref{fig:hist_global} the differences in the dust attenuations between the ``climatology'' and ``warm''  scenarios are readily apparent. However this difference only has a significant impact on link margin in the autumn and winter months of the northern hemisphere (approximately months 8 - 12 and month 1). In other months, almost all locations on Mars have margins above the threshold at zenith. However, the difference between the two scenarios is more apparent in other months when considering slant paths, as in Figs. \ref{fig:map_climatology} and \ref{fig:map_warm}.
	
	Fig. \ref{fig:hist_sites} shows the frequency of $\tau_0$ and link margin values per day throughout an entire year for the ``climatology'' and ``warm'' scenarios for the Phoenix and Curiosity landing sites and for a site in Hellas Planitia. The relatively clean atmosphere at the Phoenix landing site leads to lower $\tau_0$ and thus a higher received power, with successful link budget closure throughout the year, even in the ``warm'' scenario. In the Curiosity and Hellas Planitia sites the frequency of days with insufficient link margins at zenith are less distinct, even though the Hellas Planitia is dustier, as margin values at the Curiosity landing site are still below the threshold. The number of days (out of 360 total solar days) with link margins below 3 dB for each site are given in Table \ref{tab:total_below_margin_days}.
	
	\begin{table}[H]
		\caption{\label{tab:total_below_margin_days} Total number of days with link margins at zenith below the threshold of 3 dB.} 
		\centering
		\begin{tabular}{lllll}
			\thickhline
			\textbf{Site} & \textbf{Scenario} & \textbf{Total days} \\
			\hline
			Phoenix & Climatology & 0 \\
			Phoenix & Warm & 5 \\
			\hline
			Curiosity & Climatology & 10 \\
			Curiosity & Warm & 184 \\
			\hline
			Hellas Planitia & Climatology & 29 \\
			Hellas Planitia & Warm & 260 \\
			\thickhline
		\end{tabular}
	\end{table}
	
	\subsection{Operational scenarios}
\begin{figure*}
		\centering
		\includegraphics[width=0.85\textwidth,keepaspectratio]{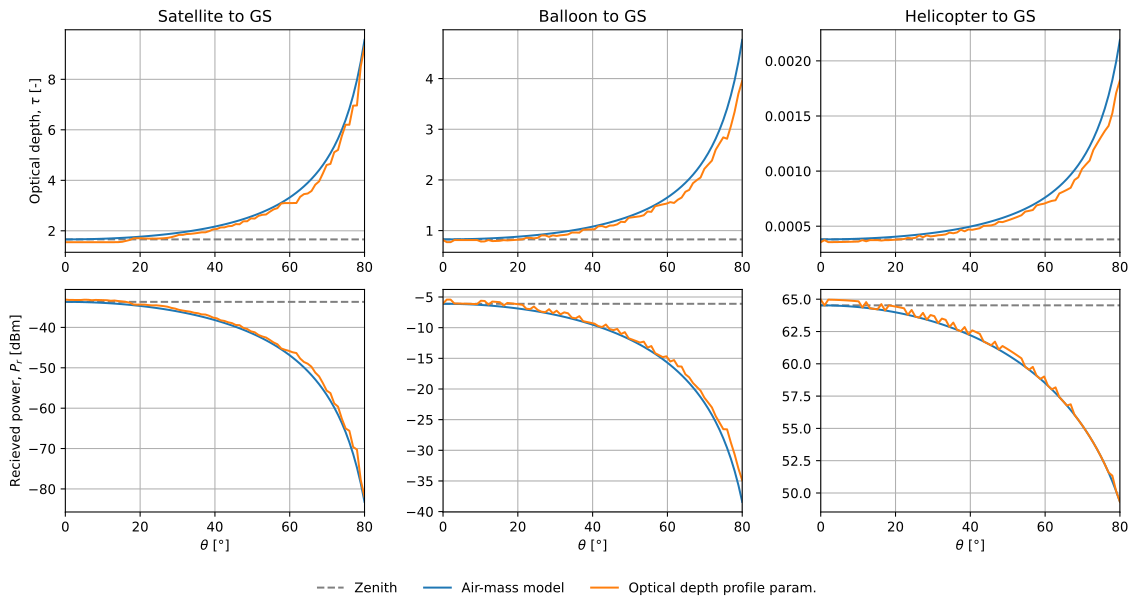}
		\caption{Optical depth (top panels) and received power (bottom panels) versus slant range for month 7, $L_s$ = 270, and local solar time 12h at the Curiosity landing site in the ``warm" scenario. Left to right panels show the satellite to ground station (GS), balloon to GS, and helicopter to GS scenarios. The values at zenith correspond to the $\tau$ and $P_{rx}$ calculations using the vertical monthly mean dust column visible optical
			depth above the surface as retrieved from the MCD ($\tau_0$; dashed line). Different curves indicate the results obtained using the air-mass model (blue) and the optical depth profile parametrisation model (orange) for calculations of $\tau$ along a slant path (see \ref{sec:slant_path_calc}).}
		\label{fig:defined_scenarios}
	\end{figure*}

	The results obtained for the scenarios described in Sect. \ref{sec:scenarios} are included in Fig. \ref{fig:defined_scenarios}. The three scenarios for satellite, balloon, and helicopter were calculated considering the ``warm'' MCD scenario for both the air-mass model (Sect. \ref{sec:slant_path_airmass}) and the optical depth profile parameterisation model (Sect. \ref{sec:slant_path_opticaldepth}). The link lengths for the satellite and balloon scenarios are extended vs. their height above the aeroid to account for the altitude of the Curiosity site (4.5 km below aeroid).
	
	Fig. \ref{fig:defined_scenarios} shows both slant optical depth ($\tau$) and received power ($P_r$) for each of the three operational scenarios with a link between the transmitter and a ground station at the Curiosity landing site. As the slant angle ($\theta$) increases, $\tau$ increases as the link travels further through the densest part of the atmosphere (which extends up to around 40 km following our parameterisation based on the dust mass mixing ratio, as can be seen in Fig. \ref{fig:mixingratio}). This leads to a consequent decrease in $P_r$. This trend is the same in all three scenarios, with only the magnitude of losses varying significantly. 
	
	The curves obtained for the optical depth profile parameterisation in Fig. \ref{fig:defined_scenarios} are less smooth compared to those for the air-mass model for all operational scenarios. This is likely because of the variations in the dust mass mixing ratio with altitude (see Fig. \ref{fig:mixingratio}), while the air-mass model depends on the smoothly varying $\cos(\theta)$.
	
	The results obtained for the link analyses made for each of the operational scenarios are provided in Tables \ref{tab:linkanalysis_sat}, \ref{tab:linkanalysis_balloon}, and \ref{tab:linkanalysis_helicopter}. These tables list the gains, losses, and outputs for the satellite, balloon, and helicopter to ground station links, respectively.
	
	\begin{table}[H]
		\caption{\label{tab:param_scenarios}Altitude above aeroid and surface as defined for each scenario.} 
		\centering
		\begin{tabular}{lllll}
			\thickhline
			Scenario & Altitude \\
			\hline
			Satellite to Ground Station & 350 km (adj. to 354.5 km)\\
			Balloon to Ground Station & 18 km (adj. to 22.5 km)\\
			Helicopter to Ground Station & 10 m\\
			\thickhline
			\multicolumn{2}{l}{* adj. = adjusted}
		\end{tabular}
	\end{table}
	
	\begin{table}[H]
		\caption{\label{tab:linkanalysis_sat} Link analysis results for the satellite to ground scenario considering a slant angle equal to zero.} 
		\centering
		\begin{tabular}{lllll}
			\thickhline
			\textbf{Parameter} & \textbf{ } & \textbf{dB | dBm} \\
			\hline
			\textbf{\textit{Gains}}\\
			\hline
			Transmitter gain &  & 77.4 \\
			Receiver gain & & 124.2 \\
			\textbf{\textit{Losses}}\\
			\hline
			Transmitter efficiency & 0.8 & \\
			Atmospheric transmission loss & & 7.2 \\
			Free space loss & & 249.2 \\
			Receiver efficiency & 0.8 & \\
			\textbf{\textit{Outputs}}\\
			\hline
			Received power &  & -33.7 \\
			Margin & & 1.8 \\
			\thickhline
		\end{tabular}
	\end{table}
	
	\begin{table}[H]
		\caption{\label{tab:linkanalysis_balloon} Link analysis results for the balloon to ground scenario considering a slant angle equal to zero.} 
		\centering
		\begin{tabular}{lllll}
			\thickhline
			\textbf{Parameter} & \textbf{ } & \textbf{dB | dBm} \\
			\hline
			\textbf{\textit{Gains}}\\
			\hline
			Transmitter gain & & 77.4 \\
			Receiver gain & & 124.2 \\
			\textbf{\textit{Losses}}\\
			\hline
			Transmitter efficiency & 0.8 & \\
			Atmospheric transmission loss & & 3.6 \\
			Free space loss & & 225.2 \\
			Receiver efficiency & 0.8 &\\
			\textbf{\textit{Outputs}}\\
			\hline
			Received power &  & -6.1 \\
			Margin & & 29.4 \\
			\thickhline
		\end{tabular}
	\end{table}
	
	\begin{table}[H]
		\caption{\label{tab:linkanalysis_helicopter} Link analysis results for helicopter to ground scenario considering a slant angle equal to zero.} 
		\centering
		\begin{tabular}{lllll}
			\thickhline
			\textbf{Parameter} & \textbf{ } & \textbf{dB | dBm} \\
			\hline
			\textbf{\textit{Gains}}\\
			\hline
			Transmitter gain & & 77.4 \\
			Receiver gain & & 124.2 \\
			\textbf{\textit{Losses}}\\
			\hline
			Transmitter efficiency & 0.8 & \\
			Atmospheric transmission loss & & 1.7e-3 \\
			Free space loss & & 158.2 \\
			Receiver efficiency & 0.8 & \\
			\textbf{\textit{Outputs}}\\
			\hline
			Received power &  & 64.5 \\
			Margin & & 100.0 \\
			\thickhline
		\end{tabular}
	\end{table}
	
	\section{Conclusions}\label{sec:conclusions}
In this paper we investigated the feasibility of FSO laser communications on Mars using a laser at 1.55 $\mu$m and assuming different dust and operational scenarios. 

Due to the negligible impact of the atmosphere itself on the total transmission, we focused on the dominant effect of the dust optical depths on the link losses. We also neglected the impact of clouds and other hydrometeors.

We presented an algorithm to estimate the vertical optical depth profile based on vertical profiles of the dust mass mixing ratio from the MCD. This model enables an alternative estimate of the optical depth expected along a slant path. Although we did not take into account the changes in optical depth expected from the changing particle size distributions with altitude, we compared our results with a simple air-mass-based estimate finding similar results. We use the direction-dependent optical depth models to analyse communication links to transmitters at three different altitudes and at different slant angles.

We conclude that global dust storms (as in the ``dust storm'' scenario of the MCD) are characterised by extremely high opacity which does not allow for optical link closure. In the case of dusty conditions (the ``warm'' scenario), the optical links are still sometimes significantly degraded making large areas of the planet unfeasible for FSO communications. During the cleaner conditions predicted by the ``climatology'' scenario, as well as during the cleaner months of the ``warm'' scenario, optical links are feasible across most of the planet.

Our results show that site selection is a key consideration for the deployment of FSO communications on Mars. Sites such as Hellas Planitia, which accumulate high dust concentrations throughout the year are particularly unsuitable for an optical link. However, even the Curiosity landing site suffers from significant degradation in link margin during dusty conditions. Therefore, the less dusty high altitude or polar regions of Mars would prove more suitable for siting a ground station. However, for the context of a future Mars mission it is important to highlight that even in these less dusty regions the link margin would still fall short of the threshold during a global dust storm. Therefore, it would be necessary to have a robust communications fallback in case of a global storm such as radio communications or local data storage for transmission when conditions improve. High altitude relay stations operating above the dust would also be interesting to consider.

In future work we intend to improve the characterization of the dust optical depth as a function of altitude and slant angle by accounting for variations in dust effective radius with altitude. Transmission modeling as a function of wavelength is also needed in order to select optimal laser wavelengths and obtain more precise optical depth data. Inclusion of a wider range of conditions at the surface and in the atmosphere (haze, fog and clouds, as well as global dust storms), and other important factors such as sky radiance levels and jitter motions would also be important to obtain a complete picture of FSO communication on Mars. Finally, an accurate understanding of the atmospheric losses on the link budget will allow for an optimization of the design requirements for the optical laser communication terminals needed at the ground station and in orbit.

	\section*{Acknowledgements}
	The MCD database is developed at Laboratoire de M\'et\'eorologie Dynamique du CNRS (Paris, France) in collaboration with LATMOS (Paris, France), the Open University (UK), the Oxford University (UK) and the Instituto de Astrof\'isica de Andalucia (Spain) with support from the European Space Agency (ESA) and the Centre National d'Etudes Spatiales (CNES).

\printbibliography	
\end{document}